\documentclass[twocolumn,showpacs,superscriptaddress,preprintnumbers,amsmath,amssymb]{revtex4}

\usepackage{epsfig}
\usepackage{psfrag}
\usepackage{float}
\usepackage{graphicx}% Include figure files
\usepackage{dcolumn}% Align table columns on decimal point
\usepackage{bm}% bold math
\usepackage{subfig}

\newcommand{\fref}[1]{Figure \ref{}}

\begin{document}
\preprint{ADP-10-4/T700}

\title{String effects and the distribution of the glue in mesons \\ at finite temperature}% Force line breaks with \\
\author{A. S. Bakry}\email[]{abakry@physics.adelaide.edu.au}
\author{D. B. Leinweber}%
\author{P. J. Moran}
\affiliation{Special Research Center for the Subatomic Structure of Matter, Department of physics, University of Adelaide, South Australia 5005, Australia.}
\author{A. Sternbeck}
\affiliation{Institut f\"ur Theoretische Physik, Universit\"at Regensburg, D-93040 Regensburg, Germany}
\affiliation{Special Research Center for the Subatomic Structure of Matter, Department of physics, University of Adelaide, South Australia 5005, Australia.}
\author{A. G. Williams}
\affiliation{Special Research Center for the Subatomic Structure of Matter, Department of physics, University of Adelaide, South Australia 5005, Australia.}

\date{6 April 2010}% It is always \today, today,
             %  but any date may be explicitly specified.
\begin{abstract}
  The distribution of the gluon action density in mesonic systems is investigated at finite temperature. The simulations are performed in quenched QCD for two temperatures below the deconfinment phase. Unlike the gluonic profiles displayed at $T=0$, the action density iso-surfaces display a prolate-spheroid like shape. The curved width profile of the flux-tube is found to be consistent with the prediction of the free Bosonic string model at large distances. 
%%PACS numbers may be entered using the \verb+\pacs{#1}+ command.
\end{abstract}

\pacs{12.38.Gc, 12.38.Lg, 12.38.Aw}% PACS, the Physics and Astronomy
                             % Classification Scheme.
\keywords{ Flux-tubes, Finite-temperature QCD, Bosonic string, Gluon flux}%Use showkeys class option if keyword
                              %display desired
\maketitle
\section{\label{sec:level1}Introduction }
  In the flux-tube model, the linearly rising potential between a pair of color sources is believed to be due to the formation of a thin gluonic flux-tube of a constant cross-section.  At high temperatures, lattice simulations on pure SU(3) gauge fields  \cite{Kac} have indicted a decrease of the effective string tension for the quark anti-quark potential with the rise of the temperature.  The QCD vacuum  structure around the sources is then expected to exhibit gluonic profiles with widths variant to the zero temperature case. The detailed geometry of the gluonic field at finite-temperature and whether it holds the constant cross-section property is an interesting topic that has not yet been explored in Lattice Quantum Chromodynamics (LQCD).   

  The low-energy dynamics of the flux-tubes in the infrared region of a confining gauge theory can be described in terms of an effective Bosonic string.  The thin flux-tube between two widely-separated color sources fluctuates like a massless string. The linearly rising part of the potential arises from the classical configuration which corresponds to the flat world sheet of the string. The quantum fluctuations of  the string lead to an universal sub-leading correction to the potential well known as the L\"uscher term \cite{luscherfr}. Lattice simulations for several gauge theories \cite{luscher,Bali,Juge:2002br,Caselle:1995fh,2004JHEP10005C} have supported the existence and the universality of the string's sub-leading effect.  At high temperatures,  the gluonic modes come into play and the effective string description of the temperature dependent quark anti-quark potential has been worked out in Refs.~\cite{deForcrand,Gao}. A further comparison with SU(3) Monte Carlo lattice data for temperatures beginning from $\rm{T} = 0.8\,\rm{T}_{c}$ \cite{Kac} has shown a good parameterizing behaviour to the string picture formula for a minimal distance of $\rm{R}\,\rm{T}=0.5$. On the other hand, there has been numerical indications that the inclusion of the higher-order string effects beyond the Gaussian approximation, e.\,g, the string's self-interaction terms in Nambu-Goto effective string action, has reproduced the correct temperature-dependent string tensions up to a temperature scale of $\rm{T} = 0.5 \,\rm{T}_{c}$ in the $\rm{3 D}$ gauge $\rm{Z}_{2}$ model \cite{caselle-2002}. 

  The string model predictions of a logarithmic broadening \cite{Luscher:1980iy} for the width of the string delocalization has also been observed in several lattice simulations corresponding to different gauge groups \cite{Bali,Pennanen:1997qm,Caselle:1995fh,Gliozzi:2010zv}. As the temperature increases, substantial deviations from the logarithmic behavior is expected, and the law of broadening turns eventually into a linear  growth for large distances before the deconfinement is reached from below \cite{allais}. Apart from the peculiar features in the laws of the broadening of the flux-tube when the temperature is raised, the string picture predicts an effect which is rather interesting from the geometrical point of view.  The width calculated at each corresponding transverse plane to the line joining the two quarks is found to differ from that at the central plane by an amount that increases with the rise of the temperature. In other words, the mesonic string picture is implying a curvature in the gluonic profile that becomes more pronounced as higher temperatures are approached. The string self-interaction with the quark line causes a noticeable difference in the delocalizations beyond the central transverse plane and these aspects remain to be ascertained in LQCD.  Moreover,  revealing the whole profile of the glue at finite temperature provides a particularly interesting source of knowledge regarding the true geometry of the flux-tube, since at finite temperature one, naturally, need not hold to any particular assumption for the shape of the gluonic source wave functions in the relevant gauge-invariant objects representing the quark states. Probing the transverse profile of the glue might even be of relevance to the modeling of ground state sector of the theory where the exact geometry of the flux-tube seems to be not yet settled \cite{heinzl-2008-78}. 

  In this paper, we investigate the distributions of the color field inside the meson at two temperatures below the deconfinement phase, $ \rm{T} \simeq 0.9\,\rm{T}_{c}$, and $ \rm{T} \simeq 0.8\,\rm{T}_{c}$. The lattice simulations are performed on the SU(3) gauge group in the quenched approximation. The field strength inside the corresponding quark system will be revealed by correlating an improved action density operator \cite{Bilson} to the mesonic state. The mesonic states are accounted for by means of Polyakov-loop correlators. Gauge-smoothing \cite{Morningstar, Moran}, in addition to a high statistics gauge-independent approach \cite{Bissey} will be employed to enhance the signal to the noise in the flux correlation function. This noise reduction approach is variant to other approaches that utilize gauge-fixing \cite{Bornyakov}. The obtained profile of the action density will then be compared to the prediction of the mesonic string models at several distances for the highest temperature near the deconfinment point $ \rm{T} \simeq 0.9\,\rm{T}_{c} $.

   The paper is organized as follows: In Sec.~II, the details of the simulation will be described. We review the predictions of the Bosonic string model for the $q \overline{q}$ potential and the width of the string fluctuations at finite temperature in Sec.~III. In Sec.~IV, we measure the quark anti-quark potential and examine the string model parameterization for various levels of gauge smoothing. In Sec.~V, the action density in mesons will be presented, the corresponding widths at several transverse planes to the tube is then measured and set in comparison with the string model predictions. Sec.~VI provides the conclusion.   
\section{Simulation Details}

\subsection{Color field measurements}
   In this investigation we have taken our measurements on 500 quenched QCD gauge-field configurations for each set of lattice parameters considered. The gauge configurations were generated using the standard Wilson gauge action on lattices with spatial volume of $36^{3}$. We chose to  perform our analysis with lattices as fine as $a = 0.1$ fm by adopting a coupling of value $\beta = 6.00$, with temporal extents of $ \rm{N}_{t} = 8 $, and $ \rm{N}_{t}= 10 $ slices, which correspond to temperatures $ \rm{T} \simeq\, 0.9 \, \rm{T}_{c}$, and $\rm{T} \simeq 0.8\, T_{c}$, respectively. The gluonic gauge configurations were generated with a pseudo-heatbath algorithm \cite{Fabricius,Kennedy} updating the corresponding three SU(2) subgroup elements \cite{Cabibbo}. Each update step consists of one heat bath and 4 micro-canonical reflections. The measurements are taken after each 2000 of updating sweeps.\\ 
The mesonic state is constructed by means of Polyakov loop correlators,
\begin{align*}
\mathcal{P}_{2Q} = \langle P(\vec{r}_{1})P^{\dagger}(\vec{r}_{2}) \rangle,
\end{align*}

\noindent where the Polyakov loop is given by,
\begin{equation}
  P(\vec{r}_{i}) = \frac{1}{3}\mbox{Tr} \left[ \prod^{N_{t}}_{n_{t=1}}U_{\mu=4}(\vec{r}_{i},n_{t}) \right],
\end{equation} 
and the vectors $\vec{r}_{i}$ define the positions of the quarks.
\noindent The measurements that characterize the color field are taken by a gauge-invariant action density operator $S(\vec{\rho},t)$ at spatial coordinate ${\vec{\rho}}$ of the three dimensional torus corresponding to an Euclidean time $t$. The measurements are repeated for each time slice and then averaged, 

\begin{eqnarray}
  S(\vec{\rho})=\frac{1}{N_{t}} \sum_{n_{t}=1}^{N_{t}}S(\vec{\rho},t).
 \label{average}
\end{eqnarray}

\noindent The action density operator is calculated via a highly-improved $\mathcal{O}(a^{4})$ three-loop improved lattice field strength tensor \cite{Bilson}.
\noindent A dimensionless scalar field that characterizes the gluonic field can be defined as,

\begin{equation}
\mathcal{C}(\vec{\rho};\vec{r}_{1},\vec{r}_{2} )= \dfrac{\langle\, \mathcal{P}_{2Q}(\vec{r}_{1},\vec{r}_{2})\rangle\, \langle\, S(\vec{\rho}) \rangle -\langle\mathcal{P}_{2Q}(\vec{r}_{1},\vec{r}_{2}) \, S(\vec{\rho})\,\rangle } {\langle\, \mathcal{P}_{2Q}(\vec{r}_{1},\vec{r}_{2})\,\rangle\, \,\langle S(\vec{\rho})\, \rangle}, 
\label{Flux}
\end{equation}

\noindent where $< ...... >$ denotes averaging over configurations and lattice symmetries, and the vector $\vec{\rho}$ refers to the spatial position of the flux probe with respect to some origin. Cluster decomposition of the operators leads to $C \rightarrow 0$ away from the quarks.
\subsection{Noise reduction}
  We make use of translational  invariance by computing the correlation on every node of the lattice, averaging the results over the volume of the three-dimensional torus, in addition to the averaging the action measurements taken at each time slice in Eq. \eqref{average}. To further improve the signal to noise ratio in the gluonic correlation function, local action reduction by smearing the gauge links has been performed on the whole 4 dimensional lattice. Since the main focus in this investigation is to resolve the nature of the flux distributions in the IR region of the theory, we have been able to show that with an appropriate level of gauge smoothing, effects on the large distance correlations can be kept minor. For each distance scale, a range of smoothing levels is seen to leave physical observables and topological structures intact. Similar techniques have been adopted in Ref.~\cite{PhysRevLett.102.032004} in the determination of the large distance $Q \overline{Q}$ force in vacuum with different levels of HYP smearing. In Sec.~IV-A, it has been shown that for a given distance scale, the measured quark anti-quark force at large distances can be left with negligible changes for a range of smoothing levels. Variant to \cite{thurner} where the Cabbibo-Marinari cooling has been employed, we have chosen to smooth the gauge field by an over-improved stout-link smearing algorithm ~\cite{Moran}. The use of this algorithm should ensure that the smoothing algorithm has a minimal effect on the topology of the gauge field \cite{Moran}. In standard stout-link smearing \cite{Morningstar}, all the links are simultaneously updated. Each sweep of update consists of a replacement of all the links by the smeared links
\begin{equation}
  \tilde{U}_\mu(x) = \mathrm{exp}(i Q_\mu(x) ) \, U_\mu(x) \,,
  \label{eqn:stoutlinksmearedlink}
\end{equation}
with
\begin{align*}
  Q_\mu(x) & = \dfrac{i}{2}(\Omega_\mu^\dagger(x) - \Omega_\mu(x)) \notag \\
  & \quad - \dfrac{i}{6}\rm{tr}(\Omega_\mu^\dagger(x) - \Omega_\mu(x)) \,,
\end{align*}
and
\begin{align*}
  \Omega_\mu(x) = \left( \sum_{\scriptstyle  \nu \ne \mu}
  \rho_{\mu\nu} \Sigma_{\mu\nu}^\dagger(x) \right) U_\mu^\dagger(x)\, ,
\end{align*}
where $\Sigma_{\mu\nu}(x)$ denotes the sum of the two staples touching $U_\mu(x)$ which reside in the $\mu-\nu$ plane.
The scheme of over-improvement requires  $\Sigma_{\mu\nu}(x)$ to be replaced by a combination of plaquette and rectangular staples.
This ratio is tuned by the parameter $\epsilon$ \cite{Moran}. In the following we use a  value of $\epsilon = -0.25$, with $\rho_\mu = \rho = 0.06 $.
We note that for a value of $\rho = 0.06$ in the over-improved stout-link algorithm is roughly equivalent in terms of UV filtering, to the standard stout-link smearing algorithm with the same $\rho=0.06$.

\section{The Bosonic String Model}
\subsection{ Temperature-dependent quark--anti-quark potential}
   The correlation function of two Polyakov loops on the lattice determines the interaction potential between the color sources,
 
  \begin{align}
\label{correlator}
  \langle P(0) \,P^{\dagger}(R) \rangle =& \int d[U] \,P(0)\,P^{\dagger}(R)\, \mathrm{exp}(-S_{w} )  ,\nonumber\\ 
                                       =& \quad\mathrm{exp}(-V(R,T)/T).
  \end{align}

\noindent $S_{w}$ is the plaquette action and $T$ is the physical temperature.
\noindent The self interactions of the glue exchanged between two color sources in QCD can result in the squeezing of the glue into a thin one dimensional string-like object. The immediate consequence of this string 
picture is that a functional form can be ascribed to the Polyakov-loop correlators. Namely, the 
partition function of the string. The correlators are expressed as a functional integrals over all the world sheet configurations swept by the string,
\begin{equation}
  \langle P(0) \,P^{\dagger}(R) \rangle= \int_{{\cal C}} [D\, X ] \,\mathrm{exp}(\,-S( X )).
\end{equation}  
\noindent The vector $X^{\mu}(\zeta_{1},\zeta_{2})$  maps the region ${\cal C}\subset R^{2}$ into $R^{4}$, with Dirichlet boundary condition $ X(\zeta_{1},\zeta_{2}=0)= X(\zeta_{1},\zeta_{2}=R)=0$, and periodic boundary condition along time direction $ X (\zeta_{1}=0,\zeta_{2}=0)= X (\zeta_{1}=L,\zeta_{2}=R)$, and $S$ is the string action and can be chosen to be proportional to the surface area, i.e. the Nambu-Goto action, 
\begin{equation}
S[X] =\sigma \int d\zeta_{1} \int d\zeta_{2} \, \sqrt{g},
\end{equation}
\noindent where $g_{\alpha \beta}$ is the two dimensional induced metric on the world sheet embedded in the background $\mathbb{R}^{4}$, 

\begin{align*} 
g_{\alpha \beta}&= \dfrac{\partial X}{\partial \zeta_{\alpha}} \cdot \dfrac{\partial X}{\partial \zeta_{\beta}},\quad (\alpha,\beta= 1,2),\\ 
g &= \det(g_{\alpha \beta}).
\end{align*}
\noindent Gauge fixing is required for the path integral (9) to be well defined with respect to Weyl 
and reparamterization invariance. The physical gauge,
$X^{1}=\zeta_{1}, X^{4}=\zeta_{2} $
would restrict the string fluctuations to transverse directions to ${\cal C}$. 
In the Quantum level, Weyl invariance is broken in 4 dimensions, however, the anomaly is known to vanish at large distances \cite{Olesen:1985pv}. The action after gauge fixing read,
\begin{align}
 S[X]& = \sigma \,\int_{0}^{L}\, d\zeta_{1} \,\int_{0}^{R}d\zeta_{2} \,(\,{1+(\partial_{\zeta_{1}} {X}_{\perp})^{2}+(\partial_{\zeta_{2}} {X}_{\perp})^{2}})^{\frac{1}{2}}.
\end{align}

\noindent Expanding the square root in powers of $\sigma R L$,
\begin{equation}
\label{path}
  S[ X ]=\sigma \,R \,L + \frac{\sigma}{2}\int_{0}^{L} d\zeta_{1} \int_{0}^{R} d\zeta_{2} (\nabla X)^{2} + .....,
\end{equation}

\noindent the action decomposes into the classical configuration 
and fluctuation part, and the string higher-order self 
interactions. A leading order approximation can be made 
by neglecting the self-interaction terms, the path integral Eq.~\eqref{path}
is then,   

\begin{equation}
  \langle P(0) \,P^{\dagger}(R) \rangle = \, e^{-\sigma R\,L}\, [\det(-\frac{1}{2}\nabla^{2})]^{-1}.
\end{equation} 

 \noindent The determinant of the Laplacian on the cylinder has been regulated using a lattice regulator in \cite{deForcrand}. The potential is obtained in closed form for a length scale comparable to the thermodynamic scale in \cite{Gao}. The effective potential is,
\begin{align}
\label{stringpotential}
V(R,T)=& \left(\sigma- \frac{\pi}{3}\, T^{2} + \frac{2}{3}\,T^{2}\, \tan^{-1}(2\,R\,T)^{-1}\right)\,R\nonumber \\ 
        & -\left(\frac{\pi}{12}-\frac{1}{6}\tan^{-1}(2\,R\,T)\right)\frac{1}{R} \nonumber\\
        & - \frac{T}{2}\, \ln(1+(2\,R\,T)^{2})+ \mu.
\end{align}
 
\noindent The limit of large string length \cite{deForcrand}, entails taking the 
temperature-dependent string tension to be,   

\begin{equation}
\label{limit}
\sigma(T)= \sigma- \frac{\pi}{3}\, T^{2}. 
\end{equation}
\noindent The free string model predicts a temperature-dependent quark anti-quark potential that is 
  featured by the existence of a logarithmic term in addition to a leading-order decrease in the 
  string tension by an amount  $ \frac{\pi}{3} \, T^{2} $. 
%%%%%%%%%%%%%%%%%%%%%%%%%%%%%%%%%%%%%%%%%%%%%%%%%%%%%%%%%%%%%%%%%%%%%%%%%%%%5 
  \subsection{Width of the Flux-tube fluctuations at finite temperature}

  \noindent The vibration modes of the string-like object renders 
  an effective width for the flux-tube. A well known prediction 
  made by L\"usher, M\"unster and Weisz \cite{Luscher:1980iy}, based on the 
  effective Bosonic string model, has shown that the mean square width 
  of the vibrating flux-tube at the center plane grows logarithmically 
  as a function of the interquark separation, R,
 \begin{equation}
 \label{Wid}  
  w^{2} \sim \frac{1}{\pi\sigma}\log(\frac{R}{\lambda})
 \end{equation}
  \noindent where $\lambda$ is an ultra-violet scale. With the increase of the 
  temperature, higher order gluonic degrees of freedom are present and the effective 
  width of the corresponding string is expected to manifest an intricate behavior  
  involving both the distance and the temperature.    
  \noindent The mean square width of the string is defined as,
  \begin{align}
  \label{operator}
   w^{2}(\xi;\tau) = & \quad \langle \, X^{2}(\xi;\tau)\,\rangle, \nonumber\\ 
                   = &\quad \dfrac{\int_{\mathcal{C}}\,[D\,X]\, X^2 \,\mathrm{exp}(-S[X])}{\int_{\mathcal{C}}[D\,X] \, \mathrm{exp}(-S[X])},
  \end{align}
  \noindent $\xi=(\xi_{1},i\xi_{2})$ is a complex parametrization of the world sheet,
  such that $\xi_{1}\in [-R/2,R/2], \xi_{2} \in [-L/2,L/2]$, with $\tau=\frac{L}{R}$ is the modular parameter of the cylinder, and $L$ is the temporal extent governing the inverse temperature. \\
  \noindent Casselle et al.~\cite{Caselle:1995fh} and Gliozi \cite{gliozzi-1994} have worked out the 
  delocalization of the string for all the planes transverse to the line joining the 
  quark pair by the corresponding Green function. This technique proceeds by removing the 
  divergence in the quadratic operator in Eq.\eqref{Wid} by the use of the Schwinger \cite{Schwinger} point-split regularization, then taking the limiting action for the Numbu-Goto model as that of the 
  corresponding Gaussian model. The quadratic operator is then the correlator   
  of the free Bosonic string theory in two dimensions,
  \begin{align}
  \langle X^{2}(\xi;\tau) \rangle = \langle&X(\xi)\cdot X(\xi+\epsilon)\rangle\nonumber, \\
                  = G&(\xi,\xi+\epsilon).\\
  \intertext{This Green function is the solution of Laplace equation on cylinder with a Dirichlet boundary condition,}
   G(\xi,\xi_{0})=&\frac{-1}{2\pi}\log\bigl| f(\xi,\xi_{0})\bigl|.\\
   \intertext{The conformal map reads \cite{allais},}
   f(\xi,\xi_{0})=&\dfrac{\theta_{1}[\pi(\xi-\xi_{0})/R;\tau]}{\theta_{2}[\pi(\xi-\overline \xi_{0})/R;\tau]},\\
   \intertext{ where the Jacobi $\theta$ functions are,} 
 \theta_{1}(\xi;\tau)=2 \sum_{n=0}^{\infty}&(-1)^{n}q^{n(n+1)+\frac{1}{4}}\sin((2n+1)\,\xi),\nonumber\\
 \theta_{2}(\xi;\tau)=2 \sum_{n=0}^{\infty}&q^{n(n+1)+\frac{1}{4}}\cos((2n+1)\xi),
 \end{align}
\noindent with the nome, $q=e^{\frac{-\pi}{2}\tau}$. The expectation value of the mean square width would then read,

\begin{equation}
\label{sol}
w^{2}(\xi_{1},\tau) = \frac{1}{2\pi\sigma}\log(\frac{R}{R_{0}})+\frac{1}{2\pi\sigma}\log\bigl|\,\dfrac{\theta_{2}(\pi\,\xi_{1}/R;\tau))} {\theta_{1}^{\prime}(0;\tau)} \bigl|.
\end{equation}

 \noindent  This expression converges for modular parameters close to $1$, and contains in addition to the 
            logarithmic divergence term a correction term that encodes the dependence 
            of the width at different transverse planes on the modular parameter of the cylinder.
            At finite temperature, this term is contributing to the width at all the planes.  
            Fig. 1 is a plot of the mean square width calculated at $\xi_{1}$ values via Eq.\eqref{sol}. 
            The plot shows the profile for several modular parameters and fixed separation 
            between the two Polyakov loops. 
                 
\begin{figure}[!hpt]
\label{profile}
\begin{center}
\includegraphics[width=8.5cm]{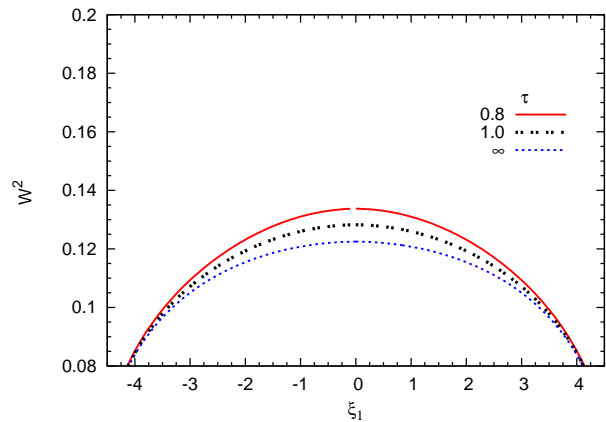}                                                             
\caption{The mean square width, Eq.\eqref{sol}, of the flux-tube evaluated at all planes $\xi_{1}$ perpendicular 
to the quark--anti-quark line. The separation distance between the pair is $R\,a^{-1}$ = 10.}
\end{center}
\end{figure}

\noindent  The string model predicts an increase in the width with the increase of the temperature. 
           The increase in the width is maximum at the central plane which is seen as an increase 
           in the curvature in the profile of the string fluctuations. At zero temperature $ L \rightarrow \infty $, 
           Eq.~\eqref{sol} converges well, and the second term in Eq.~\eqref{sol} 
           still contributes to the whole shape of the fluctuations at all planes except the middle, 
           the contribution of this term at zero temperature is,

\begin{equation} 
  \dfrac{1}{2\pi} \log \bigl|\cos(\dfrac{\pi \xi_{1}}{R})\bigl|,
\end{equation}  
           which is seen from the plot Fig. 1 to be subtle in the middle region and have more pronounced 
           effects on the width near the quark positions.

\section{Results}

\subsection{Quark--anti-quark potential}
  The technique adopted to enhance the signal to the noise in the correlation function which characterizes the gluon flux, Eq.\eqref{Flux}, involves smoothing the gauge links by the over-improved stout-link smearing algorithm described above. The whole 4-dim torus has been smeared for the consecutive levels of smearing correspond to 20, 40, 60, and 80 sweeps, forming four data sets for in each temperature. 
 \noindent The choice of the appropriate data set for the numerical evaluation of the expectation values in Eq.~\eqref{Flux} 
 at each distance scale, should be based on a compromise to simultaneously achieve two tasks, the smearing level  
 has a minimal effect on the physical observables, and a significant error reduction is gained. 
 The larger the separation distance between the quark pair, the higher smearing level required to gain good signal to noise in
 the correlations in Eq.\eqref{Flux}. However, smearing has an effect on the observables equivalent to the increase of the lattice space-time cut off, 
 and a large enough number of smearing sweeps will result in a subsequent loss of the physics on the short distance scale.
 \noindent The physical observable of direct relevance to the properties of the gluonic flux-tube is the quark anti-quark potential. For each 
 level of smearing, we numerically evaluate the quark anti-quark potential 
 and the corresponding force. At fixed temperature $T$, the Monte Carlo evaluation of 
 quark-anti quark potential at each $R$ is calculated through the Polyakov loop correlators according to,
 \begin{align}
 \label{pot}
  V(R,T)=\frac{-1}{T}\log(\langle P(0)P^{\dagger}(R) \rangle). 
 \end{align}

  \begin{figure}[htbp]
  \begin{center}
  \subfigure[] {\includegraphics[width=8cm]{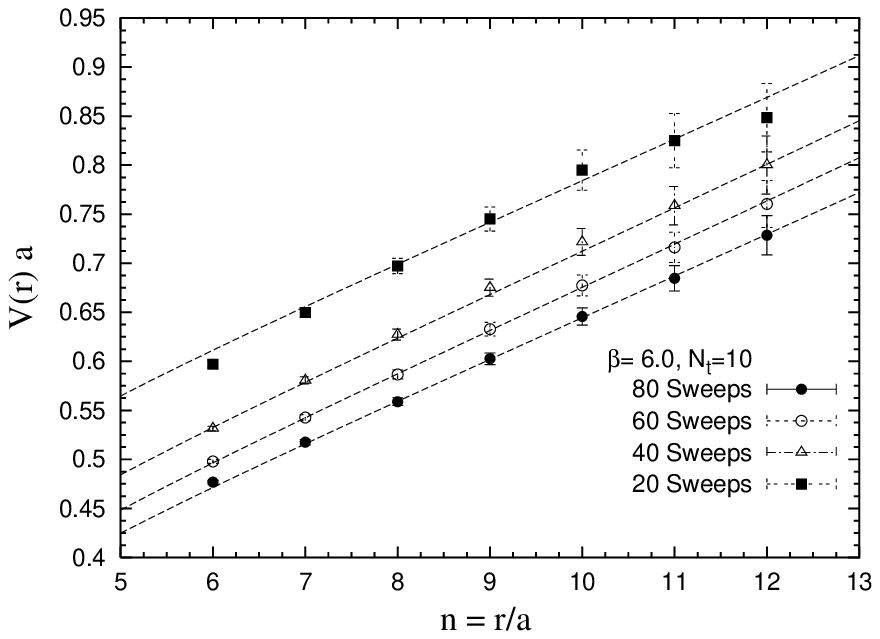} } \hfill
  \subfigure[] {\includegraphics[width=8cm]{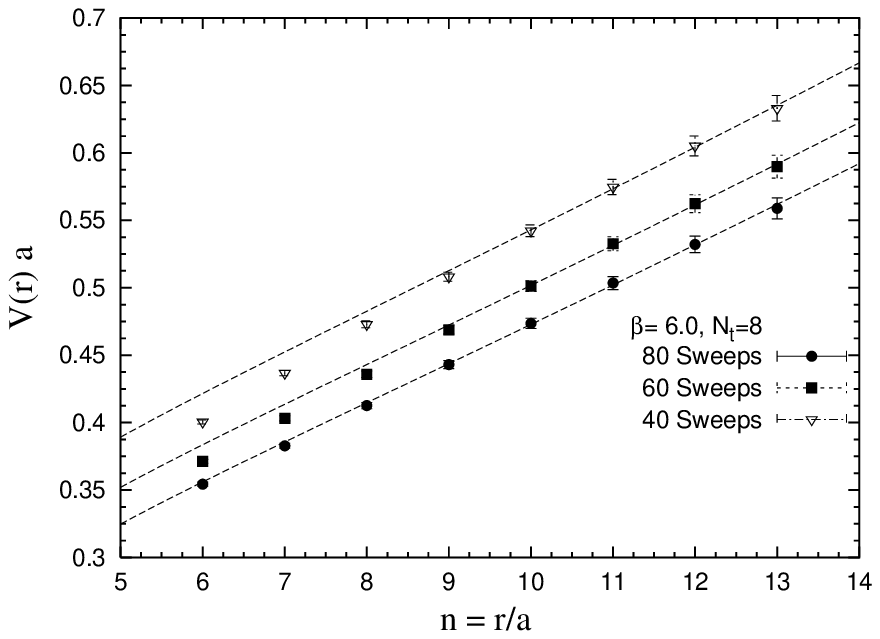}} \hfill
  \caption{ The quark-anti quark potential measured at each depicted smearing level, the lines correspond to fits of the potential obtained from the string picture of Eq.~\eqref{stringpotential} for each data set as described in the text. The upper plot is at $\rm{T} = 0.8\, \rm{T}_{c}$ while the lower plot is at $\rm{T}= 0.9\,\rm{T}_{c}$.}
  \end{center}
  \end{figure}
  \noindent The jacknife error analysis for the data shows 
  a significant decrease in the uncertainties associated with Polyakov-loop correlators on
  short distance scale when measurements are taken after 20 sweeps of smearing. 
  For large distances, a subsequent increase of 20 sweeps would provide error reduction by 
  factors of $1.3\leq x \leq 1.5$  for the corresponding distances  0.6 fm $\leq R\leq$ 1 fm. Table~I 
  summarizes the factors of error reduction for the Polyakov loop correlator after each incremental increase of 
  20 sweeps of smearing. 
\footnote{It is worth noting that by the use of a link integration method \cite{1985PhLB15177D},
  one would not expect a factor of error reduction that can be more than $x \simeq 1.1$ in the middle region. 
  For larger distances; however, link integration method would be beneficial only if supplemented by a large  
  number of measurements.}\\

\begin{table}[!hpt]
\caption{\label{Error} The error reduction factor in the Polyakov-loop correlator Eq.~\eqref{correlator} 
by the increase of the number of 20 smearing sweeps for each smeared data set.}
\begin{ruledtabular}
\begin{tabular}{cccc}
 R \slash No.Sweeps &$ 20-40 $& $40-60$ &$60-80$ \\
\hline
6  & 1.3  & 1.1 & 1.1  \\
8  & 1.4  & 1.2 & 1.2  \\
10 & 1.5  & 1.3 & 1.2   \\
\end{tabular}
\end{ruledtabular}
\end{table}
 \noindent To test the validity of the gauge smoothing approach, or equivalently, to determine the levels of smoothing for which the physics is left intact, one is tempted to set a reference scale which signifies how the smeared data would behave with respect to the string model parametrization. This approach of referencing the data to the string model is justified by the fits previously reported in Ref.\cite{Kac}, which has returned good $\chi^{2}$ 
 and shown stability to the fit range at large distances. 
 
\noindent The numerical data obtained for the quark anti-quark potential Eq.\eqref{pot}
 on every smoothed gauge configuration is fitted to the string picture $q \overline{q}$ potential 
 Eq.~\eqref{stringpotential}. The effects of smearing is expected to be more pronounced at short distances. 
 For this reason the minimal fit distances is taken as large as possible, $R > 0.7 $ fm for $ \rm{T} = 0.8 \,\rm{T}_{c}$, 
 and  $R > 0.9 $ fm for  $ \rm{T} = 0.9 \, \rm{T}_{c}$. The string tension has been taken as a fitting parameter. 
 The fits are returning good $\chi^{2}$ for all the smoothed data sets considered. 
 \noindent The quark anti-quark potential and the corresponding fits are shown in Fig.~2.
 The fits to the data show almost equal slopes for all smoothing levels. This is also manifest in Table~II, where the string tensions are measured in accord to 
 Eqs.~(11) and (12). Within the standard deviations of the measurement, 
 the string tensions for all levels of smearing are equal values. At temperature $ T = 0.9~T_{c}$, 
 our measurements for the string tension agree for all the data sets. Moreover, this value  
 is in agreement with that reported in Ref.\cite{Kac}. The factors of error reduction 
 at higher temperature at $R=1$ fm after 40 sweeps of smearing equals the corresponding one at 
 $\rm{T} = 0.8$ after 80 sweeps. The noise tends to decrease with the increase of the temperature.           
 This analysis shows that for the $q \overline{q}$ separation distances depicted in Table~II, 
 all the smoothed configurations are appropriate for revealing the gluonic field.      
 \noindent On the other hand, the data points for $R \leq 0.7 $ at $\rm{T} = 0.8$ shift up 
 towards the string model's curve with the increase of the number of smearing sweeps. The removal of short distance physics is manifest here.
 \noindent The difference in the regularization brought about by the increasing of the space time cut-off introduced by smearing, shifts the $q \overline{q}$ potential by a renormalization constant in Eq.~\eqref{pot}. To manifest the effect of smearing on the $q \overline{q}$ potential, the potential Eq.~\eqref{pot} has not been normalized. The $q\overline{q} $ force, however, can be calculated to eliminate these constant shifts. With the definition of the derivative on the lattice taken as Refs.~\cite{Sommer,luscher}, the force is computed
as,

\begin{equation}
   F(r-\frac{a}{2})= \frac{V(r)-V(r-a)}{a}.
\end{equation}
\noindent Fig. 3 shows the force calculated for all smearing 
  levels for distances up to 1.4 fm. The force from the string 
  picture, Eq.~\eqref{stringpotential}, with fit parameters 
  measured at 60 sweeps in Table~II. is illustrated. The $q \overline{q}$ force 
  calculated from the string model compares well with that data measured on
  20 sweeps of smearing for distances commencing at 0.55 fm. All data points correspond to smearing levels 
  of 20, 40 and 60 sweeps coincide, within the statistical deviations, for distances greater than 0.75 fm.
  For 20 and 40 sweeps the data convergence begins one lattice spacing earlier at R=0.65 fm.  
  The essential features of the confinement remain unchanged for the considered levels of gauge smoothing on the scale of large distances.
\begin{table}[!hpt]
\caption{\label{chi}The string tension measured on all data sets corresponding 
to various levels of link smearing. The measurements are obtained from the fits to Eq.~(11) and (12). }
\begin{ruledtabular}
\begin{tabular}{ccc}

 No. Sweeps & $\sigma a^{2}$ & Fit Range $n=R/a$ \\

\hline
\\
$ \rm{T} = 0.8 \,\rm{T}_{c} $\\
\\
   20  &0.047(3) & 8-12 \\
   40  &0.050(2) & 8-12 \\
   60  &0.0493(9)& 8-12 \\
   80  &0.0478(6)& 8-12 \\ 
\\
$ \rm{T} = 0.9\,\rm{T}_{c} $\\
\\
   40  &0.0385(8)& 10-13 \\
   60  &0.0377(9)& 10-13 \\    
   80  &0.0373(8)& 10-13 \\
\end{tabular}
\end{ruledtabular}
\end{table}

\begin{figure}[ht]
\label{force}
\begin{center}
\includegraphics[ width=8cm]{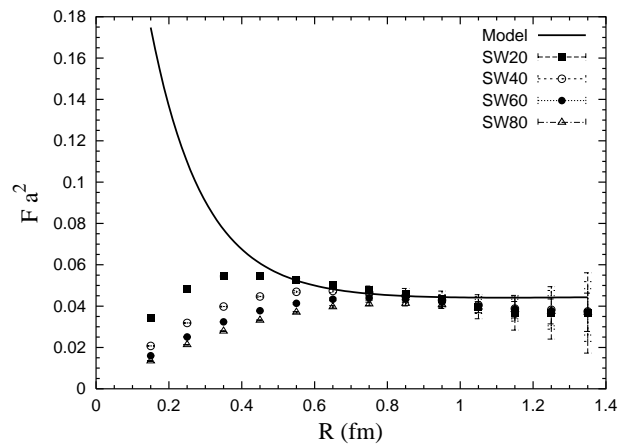}
\caption{The $q \overline{q}$ force measured for all the smearing levels up to distance of 1.4 fm, the temperature is $\rm{T} = 0.8 \rm{T}_{c}$
, $\beta =6$. The line denotes the force as predicted by the string model at finite temperature, Eq.~\eqref{stringpotential}. }
\end{center}
\end{figure}

 \noindent A combination of a large number of Monte Carlo updates  
 followed by link averaging has been performed in Ref.\cite{Kac} to evaluate
 the $q \overline{q}$ potential for a range of temperatures
 above and below the confinement phase. This involves a large number of 
 updating sweeps and measurements which makes it rather expensive in terms of the CPU time. 
 This is particularly true for the evaluation of the gluonic flux distribution, since this 
 concerns the Monte Carlo evaluation of not only the Polyakov-loop correlator, 
 but also the three-point correlation function in the numerator of Eq.\eqref{Flux}. 
 Gauge smoothing is chosen as a cheap and effective method in this case to reveal 
 the general topological features of the flux distribution which can be  
 confronted with the predictions of the string model. We have been able to show in this section the ranges of the validity of this approach, through the measurements of the physical observables that have been previously reported in \cite{Kac},i.e
 the $q\overline{q}$ potentials and the string tension.    
\subsection{Action Density}
\subsubsection{Tube profile (qualitative picture)}
\begin{figure*}[!hpt]
\begin{center}
\psfrag{Math}{$\mathcal{C}(\vec{\rho};\vec{r}_{1},\vec{r}_{2} )$}
\subfigure[$R=5~a$]{ \includegraphics[height=6.5cm, width=6.5cm]{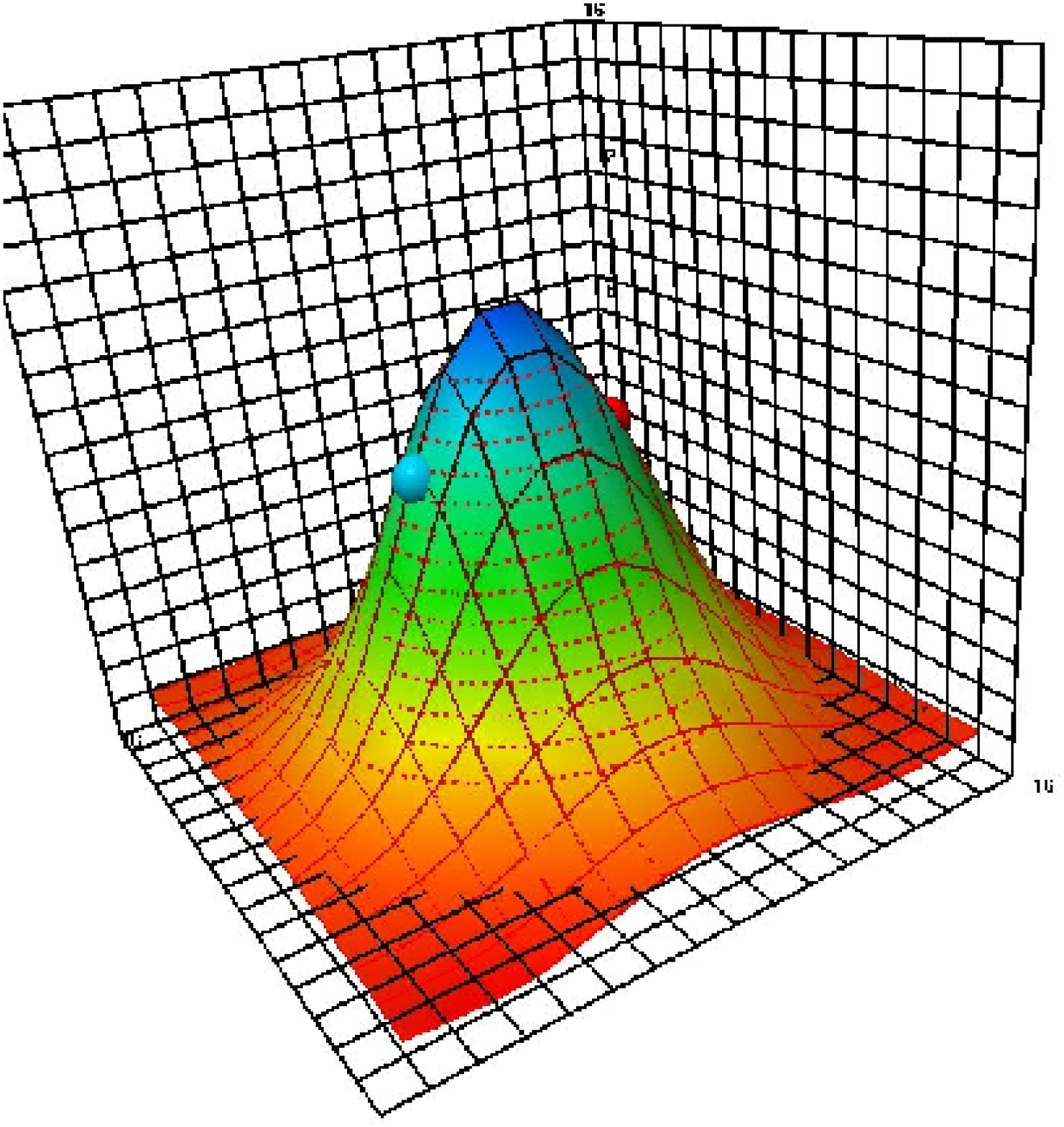} } \hspace{2cm}
\subfigure[$R=6~a$]{ \includegraphics[height=6.5cm, width=6.5cm]{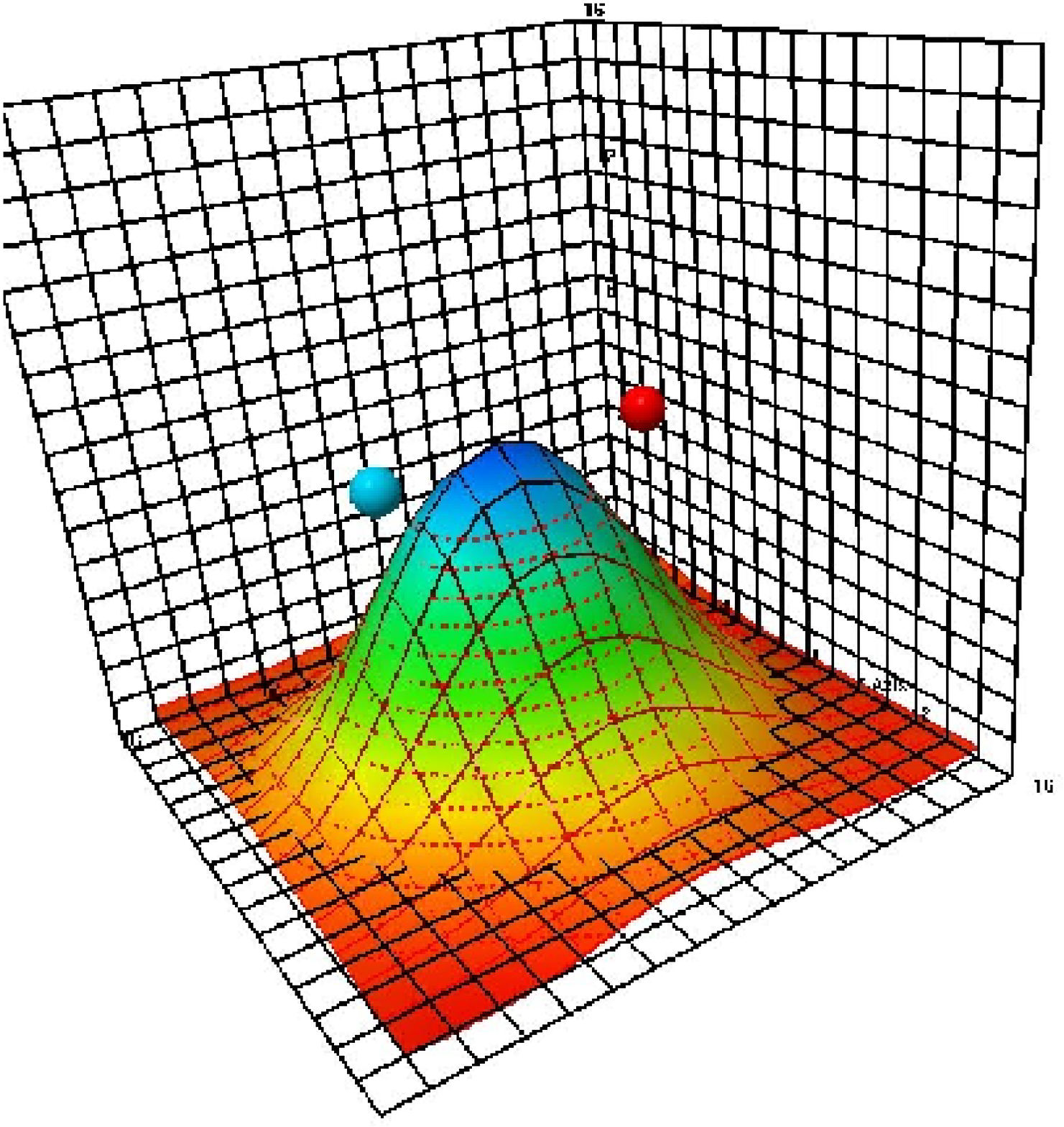}} \\
\subfigure[$R=7~a$]{\includegraphics[height=6cm, width=6cm]{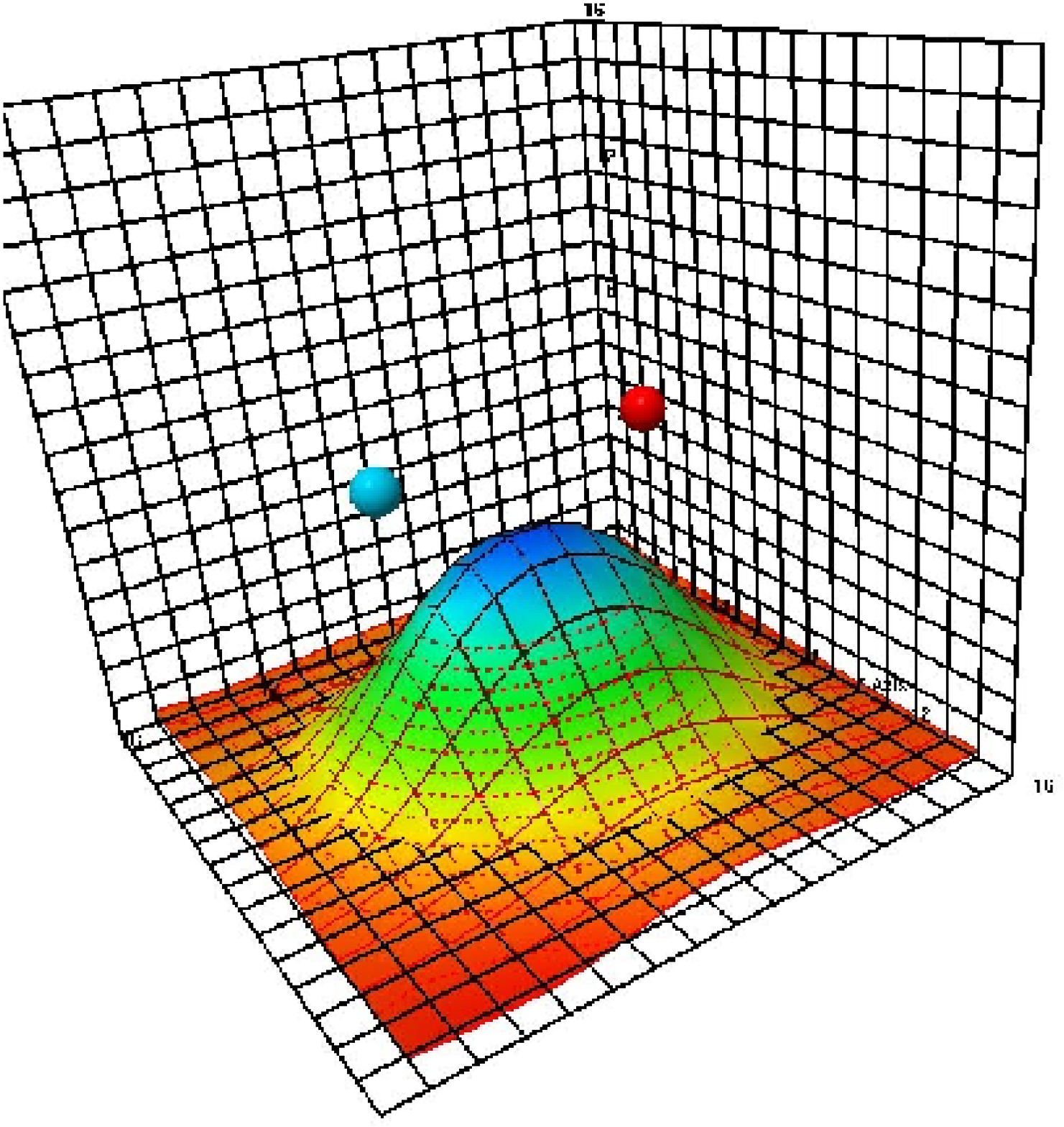} }\hspace{3cm}
\subfigure[$R=8~a$]{\includegraphics[height=6cm, width=6cm]{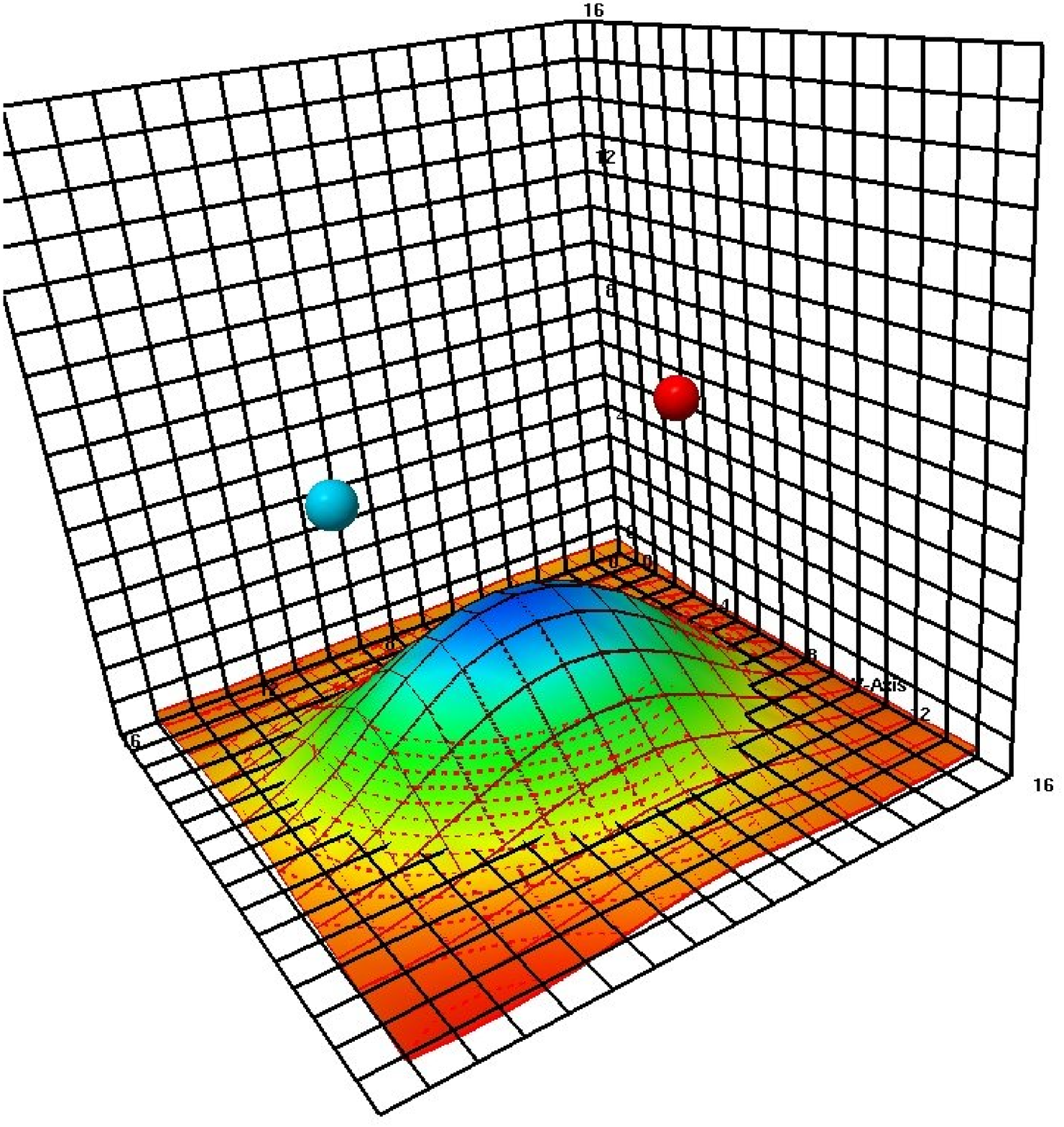} }
\caption{The flux distribution $\mathcal{C}(\vec{\rho},\vec{r}_{1},\vec{r}_{2})$ as given by the characterization Eq.~\eqref{Flux} in the plane of the quark anti-quark pair $\vec{\rho}(x,y,z=z_{0})$, for separation distances $ R $ (a)~0.5~fm,~(b)~0.6~fm,~ (c)~0.7~fm and (d)~0.8~fm at  $ \rm{T} = ~0.8 ~\rm{T}_{C}$. The spheres refer to the position of the quark and anti-quark. }
\end{center}
\end{figure*} 

 \begin{figure}[htbp]
 \begin{center}
 \subfigure[$R=9~a$]{\includegraphics[ width=8cm]{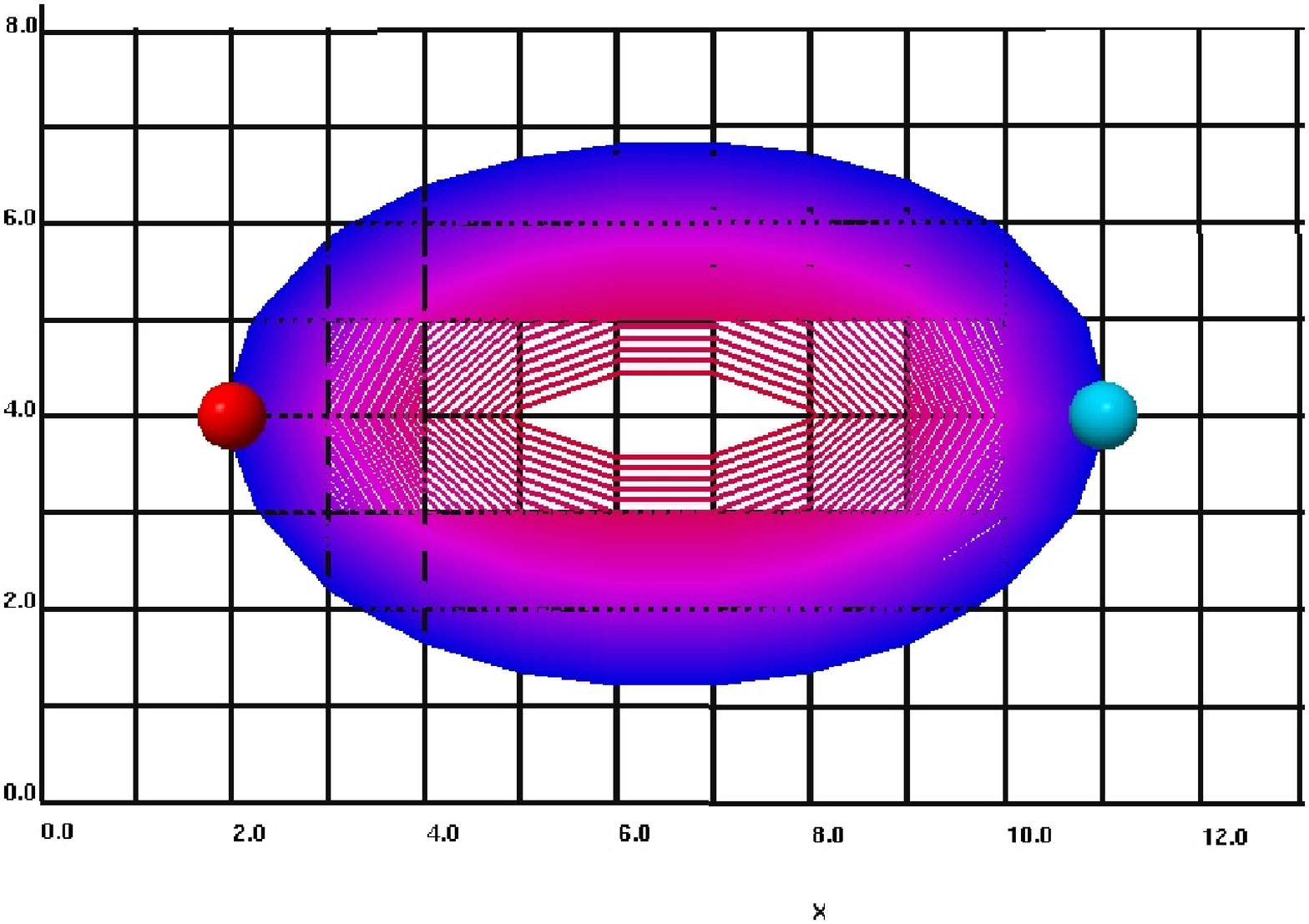} }\hfill
 \subfigure[$R=10~a$]{\includegraphics[width=8cm]{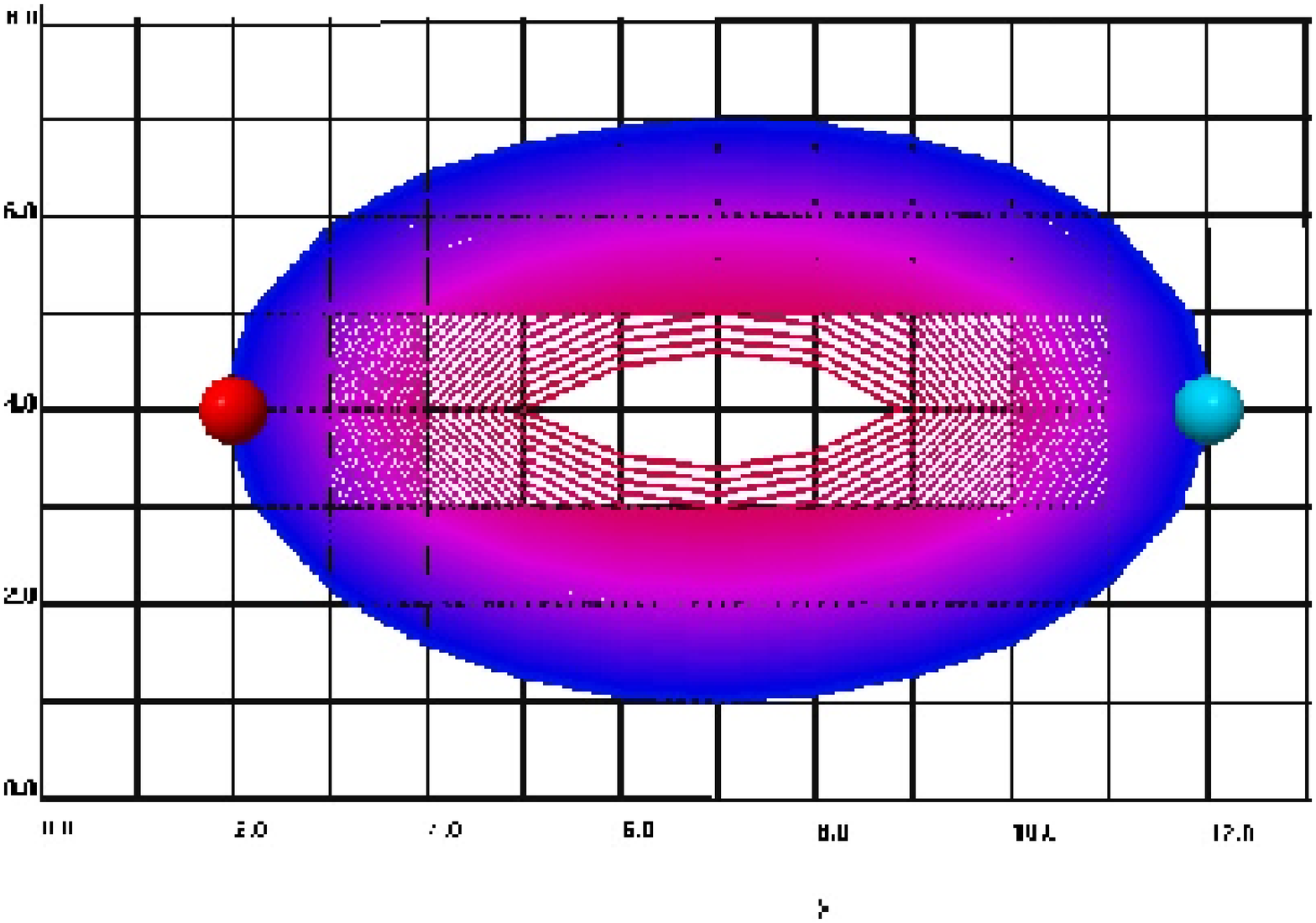} }\hfill
\caption{\label{contour}The flux contour-line distribution in the plane of the quark anti-quark pair $z_{0}$, for separation distances of (a)~0.9~fm,~(b)~1.0~fm, the spheres denote the position of the $q\overline{q}$ pair, $ \rm{T} = ~0.9 ~\rm{T}_{C}$.}
\end{center}
\end{figure}

\noindent The lattice operator which characterizes the gluonic field is usually taken as the correlation between the vacuum action-density $S(\vec{\rho},t)$, and a gauge-invariant operator representing the quark states. At finite temperature this must be a pair of Polyakov lines.
\noindent The action-density operator is calculated through an $\mathcal{O}(a^{4})$ improved lattice version of the continuum field-strength tensor. Discretization errors are reduced by combining several clover terms complemented by tadpole improvement \cite{Bilson}. 
\noindent We take our measurements with a three-loop field-strength tensor given by,
\begin{align}
    F^{\rm{Imp}}_{\mu \nu} = \sum_{i=1}^{3} w_{i}\,C_{\mu \nu}^{(i,i)} 
\label{clover}
\end{align}
 \noindent where $C^{(i,i)}$ is a combination of Wilson loop terms 
 corresponding to loops with lattice extent $i$ used to construct the clover term and $w_{i}$ are weights \cite{Bilson}. 
 \noindent The reconstructed action density, 
 \begin{align}
    S(\vec{\rho})=\beta \sum_{\mu > \nu}\, \frac{1}{2} \mathrm{Tr}(F^{\rm{Imp}}_{\mu \nu})^{2}
 \end{align}
 \noindent is accordingly measured on 20 sweeps of stout link smearing.
 This has the effect of the removal of the divergence in the action density in 
 the neighborhood of the quark positions. It is, however, very beneficial in obtaining a good signal to noise to display the flux strength. 
\begin{figure}[!hpt]
\begin{center}
\includegraphics[width=8.5cm]{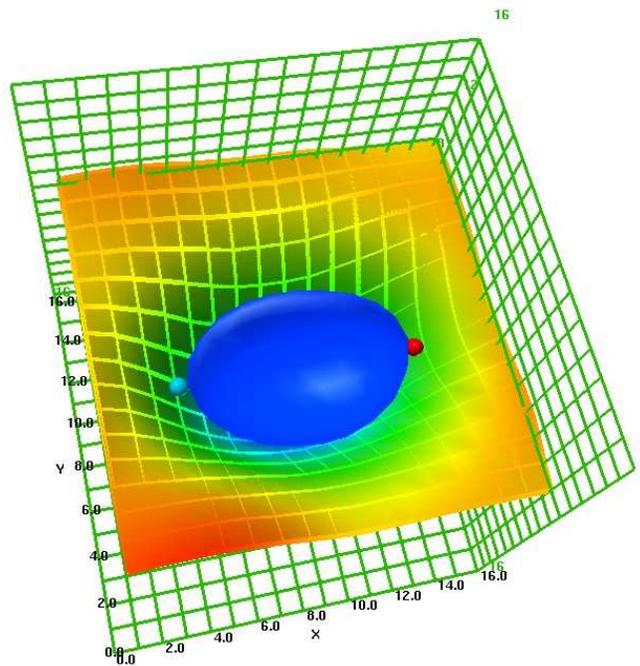} 
\caption{\label{isosurface}The flux iso-surface passing through the quarks, plotted together with a surface plot for the density distribution in the $q \overline{q}$ plane (Inverted). The measurements are taken on 80 sweeps of smearing for separation distance $R=9\,a$, and $\rm{T}\,=\,0.8\,\rm{T}_{c}$. The lattice spatial extent is $36^{3}$, $\beta=6$.}
\end{center}
\end{figure}
\noindent The correlation function Eq.~\eqref{Flux} is found $\mathcal{C}(\vec{\rho})>0$, and $\mathcal{C}\simeq 0$ away from the quark position. The scaled flux distribution in the plane of the $q \overline{q}$ pair is plotted in Fig. 3, for several $q \overline{q}$ separation distances $ \mathrm{R}=\,\mid \vec{r}_{1}-\vec{r}_{2}\mid $, at temperature  $\rm{T}\,=\,0.8\,\rm{T_{c}}$. The Polyakov-loop correlator is measured on 40 sweeps of link smoothing. The distribution shows a peak in the middle point between the $q \overline{q}$ pair at small separation distance $R=0.5 $ fm. As the two quarks are pulled apart, the distribution $\mathcal{C}(\vec{\rho})$ decreases rapidly, the peak behavior diminishes and the distribution is almost constant at $R\,=\,0.8$ fm. The qualitative description of these density plots suggests a two-dimensional Gaussian-like behavior, however, as we will see in the next section, careful measurements of the widths at each perpendicular plane to the $q \overline{q}$ line, yield different widths for large distances.

\noindent The behavior of the flux-distribution around the outer edges of the density profile does depend on finite-volume \cite{PhysRevD.47.5104}. As a by product of performing the simulations on large lattice sizes to gain high statistics in a gauge-independent manner, the two lattices employed in this investigations are being of a typical spatial size of $3.6^{3}\rm{fm}^{3}$ which do minimize the volume effect.

\noindent The curvature in the flux lines is manifesting itself as it is evident from the flux-contour plots in Fig.~\ref{contour}. The contour plots reveal the form of the flux tube just before the deconfinement phase $\rm{T}=0.9\,\rm{T_{c}}$, for $q\overline{q}$ sources separated by $R=\,0.9$~fm and $R\,=\,1$~fm, respectively. A similar plot of the action-density iso-surface at $R=\,0.9$~fm in Fig.~\ref{isosurface} displays a three-dimensional version of Fig.~\ref{contour} (prolate-spheroid like shape) for the flux-tube. This geometrical form of the density plot manifests itself at temperature $\rm{T}\,=\,0.8\,\rm{T_{c}}$ which is known to be near the end of the plateau of the QCD-phase diagram \cite{Doi2005559}.
   
\noindent It is worth noting, nevertheless, that at zero temperature, the correlation of the action density with the Wilson loop taken as a mesonic operator, does not reveal this curvature of the flux lines in the inner region between the $q \overline{q}$ pair at large separation distance \cite{Bali}. Thus we have illustrated how thermal effects show up in the action-density correlations for the first time to the best of our knowledge
\footnote{The ground-state source-wave functions in the Wilson-loop operator are trial-wave functions, and the state adopted is the one which  maximizes the overlap with the ground state, usually by smearing the string of the glue connecting the quarks \cite{heinzl-2008-78,Bali}. Moreover, the calculations of gluonic distribution is plagued by systematic errors due to biasing by the shape of the source, and the corresponding limitations imposed by the statistical fluctuations upon the Euclidean-time evolution in the loop operator \cite{Okiharu2004745,Bissey}.}.

\subsubsection{Tube profile (quantitative aspects)}

\noindent Usually studies carried out on the flux-tube laws of growth 
 focus their measurements on the central plane transverse to the $q\overline{q}$ line. At $\rm{T}=0$, it seems
 also that there is a wide belief that the tube has almost constant cross-section with a uniform energy-density profile
 for large $q \overline{q}$ separations. Nevertheless, at high temperature where the string tension is reported to decrease by a value around $10\%$  at
 $\rm{T} = 0.8\,\rm{T_{c}}$ \cite{Kac}, our calculations of the flux chromo-strength inside the meson Figs.~4-6 display a non-uniform action-density pattern around the whole $q \overline{q}$ line.

\noindent It has been conceived a long time ago  \cite{PhysRevD.11.970,thooft,Mandelstam1976245} that the QCD vacuum behaves like a dual superconductor, and the color field generated by a pair of quark sources is squeezed  into a thin string-like object dual to Abrikosov vortex by the dual Meissner effect. This squeezed flux-tube has been conjectured \cite{luscherfr} that 
 it can vibrate as a free string. At high temperature, one would expect higher modes relevant to the collective degrees of freedom of the string-like object to give rise to new interesting measurable effects, which 
 seem not only to be related to the law of the growth of the tube's width \cite{allais}, but also the width's profile it self. The string model's solution, Eq.~\eqref{sol}, informs us about how the tube would behave behind its symmetry point in the middle, together with the observed chromo-field profile, Figs.~4-6, one is tempted to investigate this string effect and establish a quantitative comparison between the model and the glue profile in QCD. This is the aim of this section.

\noindent Different possible components of the field-strength tensor in Eq.~\eqref{clover} can separately measure the chromo-electric and magnetic components of the flux. The action density, however, is related to the chromofields via $\frac{1}{2}(E^{2}-B^{2})$ and is the quantity of direct relevance to the comparison with the string fluctuations Eq.~\eqref{operator}. The width of the flux-tube may be then estimated through fitting the density distribution $\mathcal{C}(\vec{\rho})$, Eq.~\eqref{Flux}, in each selected transverse plane  $\vec{\rho}(x_{i},y,z_{0})$ to a Gaussian \cite{Caselle:1995fh,Luscher:1980iy}.

\begin{figure}[!hpt]
\begin{center}
\psfrag{C(x)}{$\mathcal{C}(\vec{\rho})$}
\includegraphics[height=8.5cm,width=6.5cm,angle=-90]{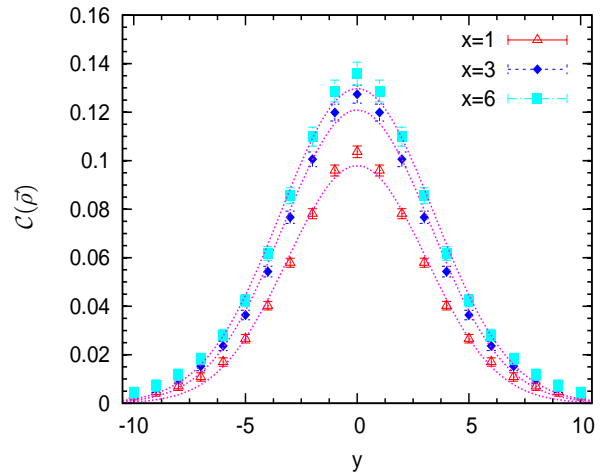} 
\caption{\label{gauss} The density distribution $ \mathcal{C} (\vec{ \rho } )$ for separation distance of $R\,=\,12$\,a, $\rm{T}\,=\,0.9\,\rm{T}_{c}$, plotted for the transverse planes $x=1$, $x=3$, and $x=6$. The lines correspond to the Gaussian fits to the density in each plane $\vec{\rho}(x_{i},y,z_{0})$.} 
\end{center}
\end{figure}

\begin{table*}[!hpt]
\caption{\label{width1}The width and the amplitude of the flux-tube at each consecutive transverse planes $x_{i}$ from the quark to the middle of the $q \overline{q}$ line. 
 The measurements for sources separation distances $R=6\,a$  to $R=10\,a$, for the temperature $\rm{T} = 0.9\,\rm{T}_{c}$.}
\begin{ruledtabular}
\begin{tabular}{cccccccccccc}
 plane&\multicolumn{2}{c}{$x=1$}&\multicolumn{2}{c}{$x=2$}&\multicolumn{2}{c}{$x=3$} & \multicolumn{2}{c}{$x=4$}  \\
  n=R/a& $A$ & $w^{2}a^{-2}$& $A$ &$w^{2}a^{-2}$& $A$ & $w^{2}a^{-2}$& $A$ & $w^{2}a^{-2}$ \\ \hline
\\
6&0.093(1)&15.6(4) &0.108(1)&15.6(4)&0.113(1)&15.7(3)\\          
7&0.099(1)&16.6(4) &0.116(1)&16.6(4)&0.125(1)&16.8(3)\\ 
8&0.101(2)&17.3(6) &0.120(2)&17.6(5)&0.131(2)&17.9(5)&0.135(2)&18.0(5)\\
9&0.102(2)&18.1(6) &0.120(2)&18.6(6)&0.132(2)&19.1(5)&0.138(2)&19.3(5)
\end{tabular}
\end{ruledtabular}
\end{table*}

\noindent The width of the tube is defined as,
\begin{equation} 
\label{widthg}
 w^{2}(x_{i})= \quad \dfrac{\int \, d^{2}\,\mathbf{\eta}\,  \mathbf{\eta}^{2} \, e^{ -( \mathbf{\eta}^{2} / w^{2} ) }} {\int \, d^{2}\,\mathbf{\eta} \,\, e^{ -(\mathbf{\eta}^{2}/w^{2})} }.
\end{equation}  

\begin{figure*}[htbp]
\label{datavsmodel1}
\begin{center}
\subfigure[$R=7~a$]{\includegraphics[width=8cm,angle=0]{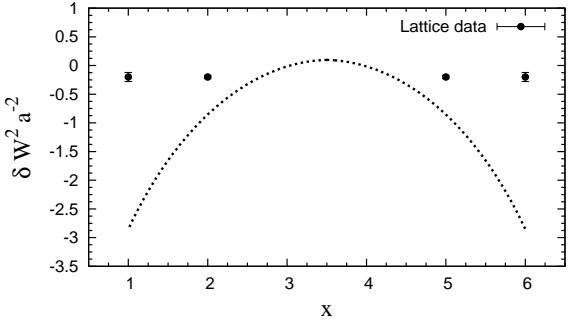} }\hfill
\subfigure[$R=8~a$]{\includegraphics[width=8cm,angle=0]{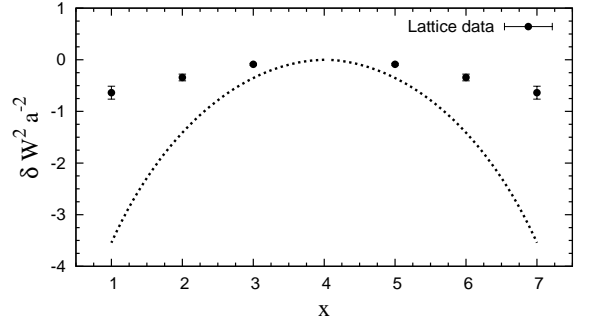} }\hfill
\subfigure[$R=9~a$]{\includegraphics[width=8cm,angle=0]{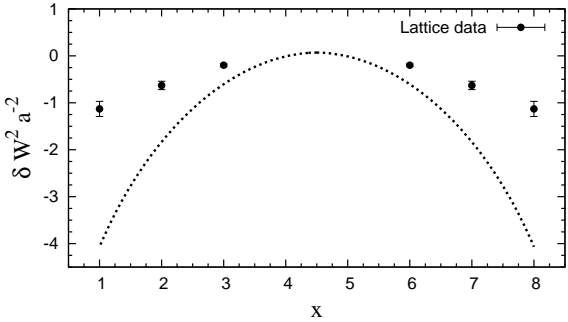} }\hfill
\subfigure[$R=10~a$]{\includegraphics[width=8cm,angle=0]{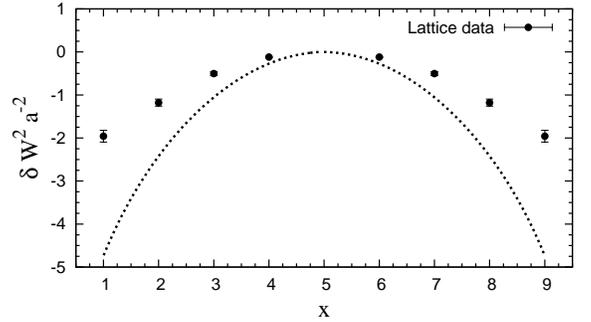}}\hfill
\caption{The width difference $\delta w^{2}= w^{2}(x_{i})-w^{2}(x_{0})$  for $q\overline{q}$ separations (a)~0.7~fm,~(b)~0.8~fm,~(c)~0.9~fm,~ and (d)~ 1~fm, ,~$\beta = ~6$, $ \rm{T} = ~0.9 ~\rm{T}_{C}$. The line denotes the width difference $\delta w^{2}$ as predicted by the string model Eq.~\eqref{sol}.}
\end{center}
\end{figure*}

\noindent $\mathbf{\eta}$ is the  set of vectors perpendicular to the $q\overline{q}$ line in the $x_{i}$ plane.

\noindent The flux calculations with Polyakov lines as a mesonic operator are well known to be distorted by statistical noise. To take reliable measurements to reveal the tube's fine structure, we choose to perform our analysis on the tube's width at the highest temperature $\rm{T}\,=\,0.9\,\rm{T}_{c}$ where the scalar field $\mathcal{C}(\vec{\rho})$ has smaller jackknife error bars, even at very large distances. To further suppress the statistical fluctuations, the number of measurements has been increased by a factor of 4. This has been done by updating the raw gauge fields and then repeating the measurements described in Sec.~II after each 70 sweeps of Monte-Carlo updates. The density distributions have been symmeterized  around all the symmetry planes of the tube, the resultant average density $\mathcal{C}(\vec{\rho})$ is fit to a Gaussian of the form $ A(x_{i})\,e^{-(y-y_{0})^{2}/w^{2}}$, with $y_{0}$  on the $q\overline{q}$ line, see e.g Fig.~7. The Gaussian fits to the data are for several transverse planes between two sources separated by a distance of $R=12\,a$. Table~\ref{width1} summarizes the measurements on both the widths $w^{2}(x_{i})$ and the amplitudes $A(x_{i})$ of the flux tube, in accord to these Gaussian fits at each transverse plane $x_{i}$ to the  $q\overline{q}$ line. The co-ordinates $ x_{i} $ are lattice co-ordinates (lattice units) and are measured from the quark position $x=0$. The uncertainties in width measurements at each transverse plane are the standard asymptotic errors in the Gaussian fits and can be assumed to be correlated. The flux-density measurements at each source separation are all taken on 40 sweeps of smearing. According to the discussion in Sec.~IV-A, this level of gauge smoothing should leave the $q\overline{q}$ potential and force with insignificant effects for $R > 0.6$ fm. 

\noindent For a fixed source separation, the measured values in Table~\ref{width1} and \ref{width2} are indicating, generally speaking, changes in the tube width along the $q \overline{q}$ line. The maximum width is measured at the tube's symmetry point in the middle. At relatively small separations $R < 0.9$ fm, the change in tube width along the planes is subtle. It is seen to vanish at $R=0.6$ fm and $R=0.7$ fm. The variation in the tube's width, however, is more pronounced at large source separation distances, see e.g Table~\ref{width2}, in general qualitative agreement with the predictions of the string picture Fig.~1. The growth in width with increasing the source separation is also maximum at the tube's center point.          

\noindent Since we focus here on comparing the tube geometry to the string profile rather than the laws of growth, we circumvent any ambiguity in measuring model's fit parameters Eq.~(\ref{sol}), by measuring the change in the width of the tube at each corresponding plane with respect to the central plane $x_{0}$, 

\begin{equation}
\delta w^{2}= w^{2}(x_{i})-w^{2}(x_{0}) ,
\end{equation}

\begin{figure*}[htbp]
\label{datavsmodel2}
\begin{center}
\subfigure[$R=11~a$]{\includegraphics[width=8cm,angle=0]{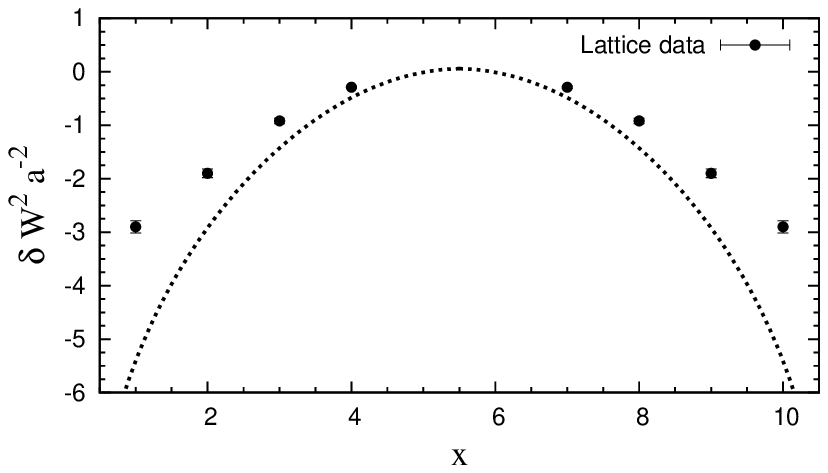}}\hfill
\subfigure[$R=12~a$]{\includegraphics[width=8cm,angle=0]{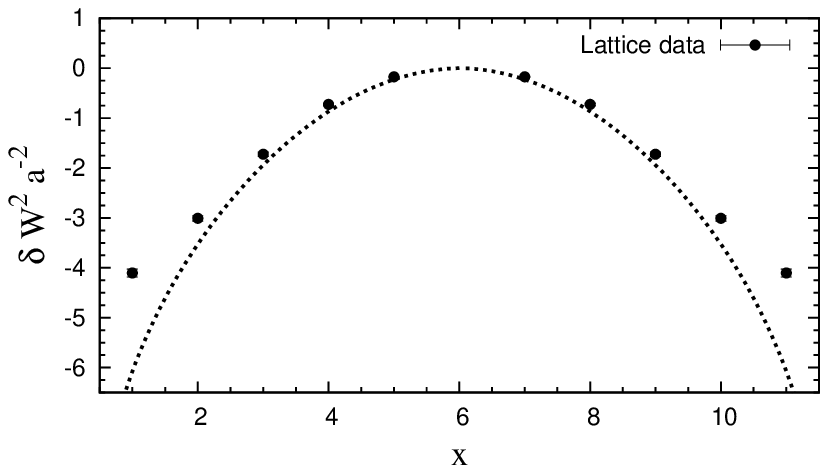}}\hfill
\subfigure[$R=13~a$]{\includegraphics[width=8cm,angle=0]{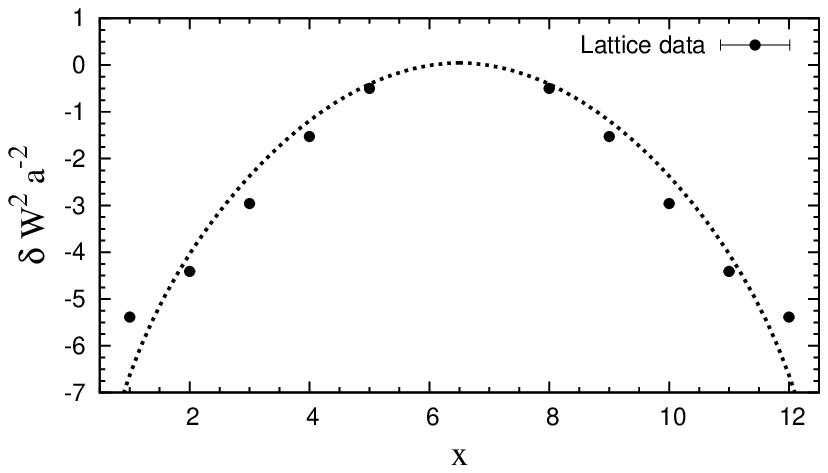}}\hfill
\subfigure[$R=14~a$]{\includegraphics[width=8cm,angle=0]{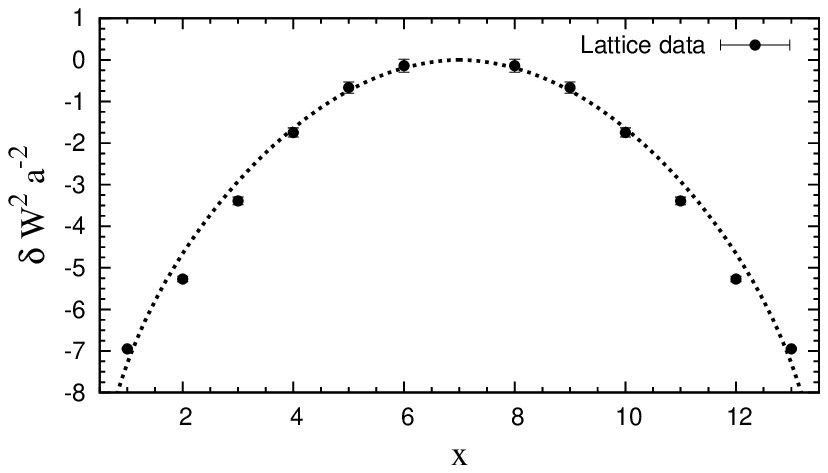}}\hfill
\caption{Similar to Fig.~8, the change in width is plotted for $q\overline{q}$ separations (a)~1.1~fm,~(b)~1.2~fm,~(c)~1.3~fm,~ and (d)~ 1.4~fm. }
\end{center}
\end{figure*}

 \begin{table*}[!hpt]
 \caption{\label{width2}Similar to Table.~\ref{width1}, the widths of the flux-tube are measured at each consecutive transverse planes $x_{i}$ from the quark to the middle of the $q \overline{q}$ line. The measurements for sources separation distances $R=10\,a$ to $R=13\,a$.}
 \begin{ruledtabular}
 \begin{tabular}{cccccccccccc}
plane &\multicolumn{1}{c}{$x=1$}&\multicolumn{1}{c}{$x=2$}&\multicolumn{1}{c}{$x=3$} & \multicolumn{1}{c}{$x=4$} &\multicolumn{1}{c}{$x=5$}&\multicolumn{1}{c}{$x=6$} \\
  n=R/a& $w^{2}a^{-2}$&$w^{2}a^{-2}$& $w^{2}a^{-2}$& $w^{2}a^{-2}$& $w^{2}a^{-2}$ & $w^{2}a^{-2}$  \\ \hline
\\
10 & 18.7(7) & 19.4(6)& 20.1(6)& 20.5(6)& 20.6(6)\\
11 & 19.0(5) & 20.0(6)& 21.0(6)& 21.6(6)& 21.9(6)\\
12 & 19.2(7) & 20.3(6)& 21.6(6)& 22.6(7)& 23.2(6)& 23.3(6)\\
13 & 19.3(6) & 20.3(5)& 21.7(7)& 23.1(7)& 24.1(7)& 24.6(7)
 \end{tabular} 
 \end{ruledtabular}
 \end{table*}
\noindent this can provide a measure on how rounded or squeezed the flux tube would be compared to the width of the string fluctuations. Fig.~8 shows the change in the tube width calculated for separation distance $R= 0.7$ fm to $R=0.9$, with uncertainties taken assuming the standard errors in the Gaussian fits are correlated, i.e. $\mid e(x_{i})-e(x_{0})\mid $. In contrast with the predictions of the string model, the tube has almost constant width at $R=0.7\,a$, the measured changes in width at the plane $x=1$ deviate from the model at $R=0.8$ fm and $R=0.9$ fm by large values of 82\%  and 70\%, respectively. The deviations decrease as the sources are pulled apart to 54\% at $R=10$, $38\%$ at $R=12$, and good agreement between both profiles is reached at $R=14$ as can be seen in Fig.~9. The change of the width measured at the inner-transverse planes, however, agrees with the model at shorter distances, $R=12$ for the plane $x=2$ and $x=3$. In general, the four plots in Fig.~9 show significant improvement with respect to the model predictions compared to the four plots at shorter distances in Fig.~8.      
\noindent  The flux-tube shows a constant cross section for $ R = 0.6, 0.7$ fm in disagreement with he string picture. At distances $0.7\,\rm{fm} < R < 1.0\,\rm{fm}$, the LGT gluonic distribution profile is, geometrically speaking, more squeezed than the free-string picture would imply. As the sources are pulled further apart, the disagreement decreases gradually and the profiles of the glue and the string both compare well for sources separations $R \gtrsim 13$ fm. 

\noindent The thermal effects are manifest in the gluonic profile, giving rise to non-uniform widths. The string picture can parameterize these profiles only at large distances. At short distances on the other hand, the free-string picture does not seem to model the gluonic interactions on the scale of short distances which may become even more relevant in the thermal regime.  
\subsubsection{Tube growth in width} 
   
\begin{figure*}[htbp]
\label{datavsmodel3}
\begin{center}
\subfigure[$x=1$]{\includegraphics[width=8cm,angle=0]{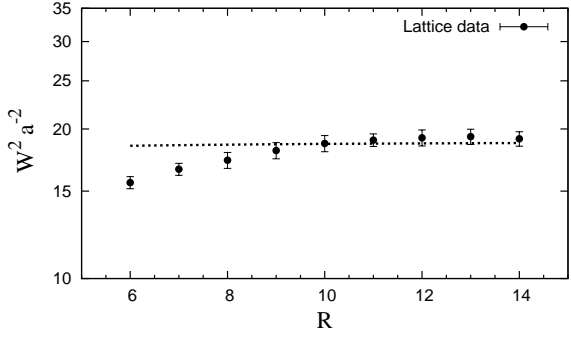}}\hfill
\subfigure[$x=2$]{\includegraphics[width=8cm,angle=0]{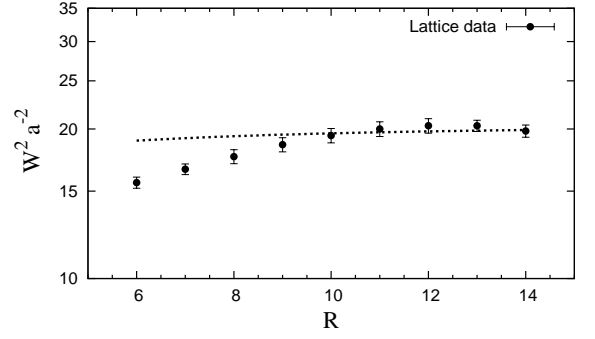}}\hfill
\subfigure[$x=3$]{\includegraphics[width=8cm,angle=0]{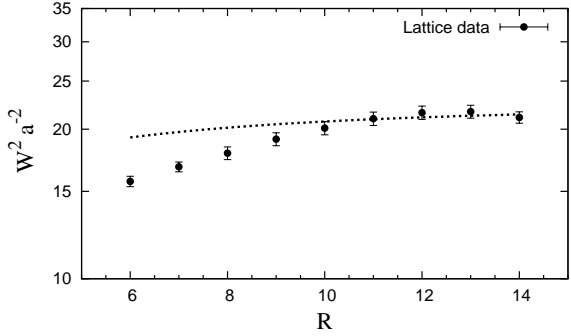}}\hfill
\subfigure[$x=4$]{\includegraphics[width=8cm,angle=0]{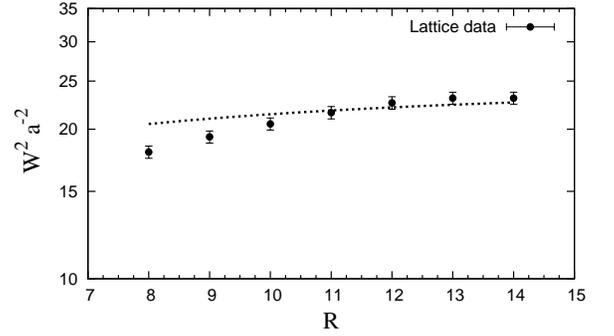}}\hfill
\caption{The width  $ w^{2}(x_{i}) $  for $q\overline{q}$ separations $R=6\,a$ to $R=13\,a$ at four consecutive planes (a)~$x=1$,~(b)~$x=2$,~(c)~$x=3$, and (d)~$x=4$. ~$\beta = ~6$ , ~$ \rm{T} = ~0.9 ~\rm{T}_{C}$ . The line denotes the string model, Eq.~\eqref{sol}, fit to the data as described in the text.}
\end{center}
\end{figure*}

\begin{table*}
\caption{\label{width3}The resultant measurements of the scale $R_{0}$ for the first four consecutive transverse planes $x_{i}$ in accord to the fits of the tube width to the string model formula Eq.~\ref{sol}. The values of the fit parameter and the corresponding $\chi^{2}_{\rm{dof}}$ are presented for variety of fit ranges.}
\begin{center}
\begin{ruledtabular}
\begin{tabular}{cccccccccccc}
 Plane&\multicolumn{2}{c}{$x=1$}&\multicolumn{2}{c}{$x=2$}&\multicolumn{2}{c}{$x=3$} & \multicolumn{2}{c}{$x=4$}& \\
  Fit range&$R_{0}$ & $\chi^{2}_{\rm{dof}}$& $R_{0}$ &$\chi^{2}_{\rm{dof}}$& $R_{0}$ &$\chi^{2}_{\rm{dof}}$&$R_{0}$&$\chi^{2}_{\rm{dof}}$& \\ \hline
\\
 6-10&0.0155(3) &4.3&0.028(5)&5.9&0.033(5)&5.5\\        
 6-13&0.0126(2)&5.9&0.022(4)&8.9&0.027(4)&9.2\\          
 7-13&0.0108(1)&3.1&0.019(3)&5.7&0.023(4)&7.1\\
 8-13&0.0093(8)&1.2&0.017(2)&2.4&0.019(3)&3.1&0.020(3)&4.7\\
 9-13&0.0086(5)&0.4&0.014(1)&0.9&0.017(2)&1.7&0.018(3)&2.9\\
 10-13&0.0081(2)&0.1&0.0129(6)&0.3&0.015(1)&0.7 & 0.015(1)& 1.2 \\ 
\end{tabular}
\end{ruledtabular}
\end{center}
\end{table*}
   The measured values in Tables~\ref{width1} and \ref{width2} indicate a growth in the tube's mean square width at all transverse planes $x_{i}$ as the color sources are pulled apart. The growth in flux-tube width at each selected transverse plane can be compared to the corresponding growth in the string fluctuation Eq.~\eqref{sol}, this comparison can be performed by the fitting of formula Eq.~\eqref{sol} to the tube measured widths. Table~\ref{width3} summarizes the resultant measurements of the fit parameter and the corresponding $\chi^{2}_{\rm{dof}}$ at four consecutive transverse planes $x=1$ to $x=4$. The fits show strong dependency on the fit range if the points at small sources separations are included. The highest value of $\chi^{2}$ is returned when fits include the whole range of sources separations, i.e.~ $R=6$ to $R=13$. With the first four points excluded from the fit, the returned $\chi^{2}_{\rm{dof}}$ is smaller indicating that the data points at large source separation are better parameterized by the string model formula. The value of the $\chi_{\rm{dof}}^{2}$ gradually decreases as we exclude points at short distance separations, and stability in the fit is reached for widths measured for the plane $ x = 1$ at sources separations $ R > 0.7$ fm, and at $ R > 0.8 $ fm for the plane $x = 2 $. The fits are returning good $\chi_{\rm{dof}}^{2}$ values for fits at the planes $x=3, 4$ for sources separations $R > 0.9$ fm. In the regions where the fits are returning good $\chi^{2}_{\rm{dof}}$, the values of the fit parameters are almost equal for the planes in the middle, $x = 2, 3, 4$. However, at the closet plane to the sources, $x=1$, the value of the returned parameter, unsurprisingly, deviates from the corresponding one at other planes. This is a manifestation for the above observed deviations in the change in tube widths at this plane compared to the central plane Figs.~8 and 9. 
 
\noindent Fig.~10 shows data points and the corresponding best fits to the string model at each plane, the string model at finite temperature poorly describes the lattice data at short distances. The plots depict the fact that the flux tube observed in LGT has a more suppressed profile than the fluctuations of the free string would imply at short distances. On the other hand, the predicted growth of the flux-tube diameter is manifest in the lattice data.    

\section{Conclusion}
   The gluonic-distribution inside the meson has been revealed. The Monte-Carlo simulations have been performed on SU(3) gauge-group for temperatures $\rm{T}\,\simeq \,0.8\,\rm{T}_{c}$ and $\rm{T}\,\simeq\,0.9\,\rm{T}_{c}$. Noise reduction has been achieved by a gauge-independent high statistics approach, in addition to the employment of adequate levels of gauge smoothing that preserve the relevant physics at large distances. This method is variant to noise-reduction by Abelian gauge-fixing. The flux tube, characterized as a correlation between the action-density and the mesonic operator (Polyakov-lines), has been displayed up to distances of 1.4 fm. The flux iso-lines and iso-surfaces display a curved profile along the tube. The profile is showing a non-uniform action-density pattern unlike that observed using Wilson's loop as a mesonic operator at $\rm{T}=0$. The flux tube width profile is compared to the corresponding mean-square width of the free Bosonic string fluctuations at all planes between the color sources. For source separation distances $R>0.7$ fm, measurements of the tube cross-section at each selected transverse plane show a non-constant width for the tube with maximum width at the symmetry point of the tube. At small $q\overline{q}$ source separation $0.7\,\rm{fm}<\,R\,<\,1.1\,\rm{fm}$, the tube is seen to yield a more compact (squeezed) form than the string model would predict. The deviations of the tube width profile from the corresponding string profile decrease gradually as source separation increases and the profiles begin to compare well at $R \backsimeq 1.3\,\rm{fm}$. 
\noindent The gluonic profiles displayed in this investigation are geometric manifestations of thermal effects on the $q \overline{q}$ potential (the measured decrease in the string tension). Moreover, the squeezed gluonic profile in comparison to the rounded string fluctuations provides also a geometrical interpretation for the deviations of the predicted string tension based on the free string picture from the corresponding lattice results \cite{Kac}. 

This study is motivating further investigations of the energy-density and chromo-electromagnetic distributions with methodological improvements that minimize the number of smearing sweeps and increase the number of measurements. It would also be interesting to confront these profiles with the Bosonic string profiles in the context of string's self-interactions. The string's geometrical effects (curved profiles) ought to be addressed in other gauge groups. Work is progressing in these directions, in addition to the detailed investigation of the thermal hadronic gluonic distributions by straight forward generalizations of the calculations presented here to Baryons \cite{AIP,Tp}.  
%\newpage %Just because of unusual number of tables stacked at end

\begin{acknowledgments}
Thanks to Tom Cohen for providing useful comments. We also thank Dimitri Diankov for encouragement to conduct the investigation.  
This work has been done using the super computing resources from the NCI National Facility and eResearch SA. The 3-D and 2-D realizations have been rendered in the eResearch SA Visualization Lab. 
\end{acknowledgments}

\bibliography{Stringeff}% Produces the bibliography via BibTeX.

\end{document}